\documentclass[reprint,
showpacs,amsmath,amssymb,aps,prb]{revtex4-1}

\usepackage{graphicx}
\usepackage{dcolumn}
\usepackage{bm}
\begin{document}

\preprint{APS/123-QED}

\title{Full-counting statistics of transient energy current in mesoscopic systems}

\author{Zhizhou Yu}
\author{Gao-Min Tang}
\author{Jian Wang}
\email{jianwang@hku.hk}
\affiliation{Department of Physics and the Center of Theoretical and Computational Physics, The University of Hong Kong, Hong Kong, China and The University of Hong Kong Shenzhen Institute of Research and Innovation, Shenzhen, China}

\date{\today}

\begin{abstract}
We investigate the full-counting statistics (FCS) of energy flow carried by electrons in the transient regime. Based on two measurement scheme we formulate a non-equilibrium Keldysh Green's function theory to compute the generating function for FCS of energy transport. Specifically, we express the generating function using the path integral along Keldysh contour and obtain exact solution of the generating function using the Grassmann algebra. With this formalism, we calculate the transient energy current and higher order cumulants for both single and double quantum dot (QD) systems in the transient regime. To examine finite bandwidth effect of leads to FCS of energy transport, we have used an exact solvable model with a Lorentizian linewidth where all non-equilibrium Green's functions can be solved exactly in the time domain. It is found that the transient energy current exhibits damped oscillatory behavior. For the single quantum dot system the frequency of oscillation is independent of bandwidth of the leads while the decay rate of the oscillation amplitude is determined by the lifetime of resonant state which increases as the bandwidth decreases. At short times, a universal scaling of maximum amplitude of normalized cumulants is identified for the single QD system. For the double QD system, the damped oscillation of energy current is dominated by Rabi oscillation with frequency approximately proportional to the coupling constant between two quantum dots. In general, the transient energy current increases when the coupling between two QDs is stronger. However, when the interdot coupling is larger than half of the external bias the transient energy current is suppressed significantly. All these results can be understood analytically.
\end{abstract}

\maketitle

\section{Introduction}\label{sec1}

With the rapid development of nanotechnology, the transport of single electron can now be studied in electronic devices \cite{Lu,Bylander,Fuji}, leading to renewal interest on statistical distribution of electrons and the energy it carries. Full-counting statistics is a new methodology to characterize full probability distribution of electron and energy transport by calculating the corresponding generating function \cite{Belzig,Bagrets,Pilgram,Gogolin,Scho,Saito,Flindt,Urban,T1}. Experimentally the real-time counting of electrons has been carried out in a quantum dot (QD) which direct measures the distribution function of current fluctuations \cite{Bylander,Gust,Flindt2}. In the experiment, the measurement of higher order \textit{cumulants} up to 15th has also been reported in quantum point contact systems.\cite{Flindt2} Such a stochastic process shows universal fluctuation for noise and higher order moments which is a common feature in mesoscopic systems. Full-counting statistics offers a superior way to study the noise and those all higher order correlations. Furthermore, quantum entanglement and quantum information have been also related to quantum transport in terms of full-counting statistics\cite{Klich,Song,Les}.

One of the important issues in the non-equilibrium transport process is the energy transport which gives information on how energy is dissipated and its correlations for any working electronic devices. Recently, the energy transport through the one-dimensional system, such as the trapped ion chains, has been measured experimentally \cite{Ramm}. The energy dissipation and fluctuation can be characterized by the energy current, which can be investigated by Landauer-B\"{u}ttiker type of formalism theoretically in the dc transport for non-interaction electrons \cite{But,sivan,kearney}. The energy current $I^E_\alpha$ is also related to the heat current $I^h_\alpha$ by $I^h_\alpha=I^E_\alpha-\mu I_\alpha$ where $\mu$ is the chemical potential and $I_\alpha$ is the particle current. It is known that the Joule heating $J$ due to the leads is related to the heat current by $\sum_\alpha I^h_\alpha=J$. The heat current driven by external bias and temperature gradient plays a central role in studying the efficiency of nano-scaled heat engine\cite{engine1}. Recently ac heat current, its shot noise, and relaxation resistance have been investigated in mesoscopic systems\cite{sanchez1,sanchez2,mosk1,mosk2,jchen}. Using the non-equilibrium Green's function transient heat current has been studied in mesoscopic systems\cite{Michelini}. Moreover, A first-principles calculation for transient heat current through molecular devices has also been carried out\cite{Yu}. It would be interesting to further study the FCS of energy transport in ac regime.

We note that FCS of energy transport of phonon has been studied extensively. An exact formula for cumulant generating function of heat transfer has been derived in harmonic networks to study non-equilibrium fluctuations \cite{SD1,SD2}. Moreover, energy fluctuations in a driven quantum resonator has also been studied by full-counting statistics \cite{C1,C2}. Using the phonon non-equilibrium Green's function, the generating function has been obtained for phonon transport in the transient regime as well as steady states\cite{W1,W2,W3}. Various cumulants of thermal current and entropy production have been studied numerically\cite{W1,W3,W4}. So far most of investigations of FCS of energy transfer focus on the phonon transport. However, less attention has been paid on FCS of the electron energy transfer\cite{foot5}. It is the purpose of this paper to address this issue.

In this paper, we develop a Keldysh non-equilibrium Green's function (NEGF) theory to study FCS of transferred energy in transient regime. Based on two measurement scheme we derive the expression of generating functions for FCS in terms of non-equilibrium Green's functions. This allows us to calculate $n$th cumulant $C_n$ of transferred energy of mesoscopic systems in the transient regime as well as in long-time limit. We then apply our formalism to investigate the energy transport in the transient regime for both single and double QD systems. To study the finite bandwidth effect on the cumulants of transferred energy, an exact solvable model is used with Lorentzian linewidth so that the non-equilibrium Green's function can be obtained exactly. As expected, the cumulants of transferred energy show linear characteristics in time in the long-time limit for both systems. For the single QD, the transient energy current exhibits damped oscillatory behavior. The oscillation frequency is found to be independent of bandwidth of the leads and the decay of oscillation amplitude is proportional to the lifetime of resonant state of QD which decreases as the bandwidth increases. At short times, the maximum amplitude $M_n$ of the normalized $n$th cumulant $C_n(t)/C_1(t)$ show a universal behavior for the single QD system. Specifically, we find $M_{2k} = a_1 e^{\kappa k}$ and $M_{2k+1} = a_2 e^{\kappa k}$ for different system parameters where $a_1$ and $a_2$ are non-universal constants. The universal slope $\kappa$ is found to be close to 3. For the double QD system, we find that the transient energy current shows the damped Rabi oscillations with frequency approximately proportional to the interdot coupling constant $v$ between two QDs. A threshold of interdot coupling $v_c$ is found below which transient energy current increases with the increase of $v$ while for $v>v_c$ the transient current is suppressed significantly. These interesting results can be understood analytically.

This paper is organized as follows. In Sec.~\ref{sec2}, the formalism of generating function for studying full-counting statistics of transferred energy in the transient regime is first presented. In Sec.~\ref{sec3}, we apply the formalism obtained to both single and double QD systems and show numerical results of various cumulants, transient energy current and the corresponding higher order cumulants. Finally, the discussion and conclusion are given in Sec.~\ref{sec4}.

\section{Theoretical Formalism}\label{sec2}
To study full-counting statistics of energy transport, we need to obtain the probability distribution $P(\Delta \epsilon, t)$ of the transferred energy carried by electrons $\Delta \epsilon = \epsilon_t - \epsilon_0$ between an initial time $t_0$ (for simplicity, we set $t_0=0$) and a later time $t$ which can be calculated from two-time quantum measurement. Denoting $\epsilon$ the eigenvalue of the Hamiltonian of the left lead $H_L$ where we measure the energy flow. Taking measurement at $t$ gives $\epsilon_t$ which is a stochastic variable. The generating function $Z(\lambda, t)$ with the counting field $\lambda$ can be obtained by the Fourier transformation of the probability distribution as \cite{Scho},
\begin{equation}\label{Z}
  Z(\lambda,t) \equiv \langle e^{i\lambda \Delta \epsilon}\rangle = \sum_{\Delta \epsilon}P(\Delta \epsilon, t) e^{i\lambda \Delta \epsilon}.
\end{equation}
The $j$th cumulant of transferred energy $\langle \langle(\Delta \epsilon)^j\rangle\rangle$ is defined by,
\begin{equation} \label{jth}
  \langle\langle (\Delta \epsilon)^j \rangle\rangle = \frac{\partial^j \ln Z(\lambda)}{\partial (i\lambda)^j} \bigg{|} _{\lambda=0}.
\end{equation}

We now derive the generating function using NEGF theory for a general QD system coupled with two semi-infinite leads in the transient regime. For this purpose, we assume that the couplings between the QD and leads are turned on at $t=0$. The Hamiltonian of the whole system can be written as,
\begin{equation}\label{ham}
  H = \sum_{k\alpha} \epsilon_{k\alpha} c^\dag_{k\alpha} c_{k\alpha} + \sum_n \epsilon_n d_n^\dag d_n +  \sum_{k\alpha n} \Big( t_{k\alpha n}c^\dag_{k\alpha}d_n  + \mathrm{h.c.} \Big),
\end{equation}
where, $c^\dag (c)$ and  $d^\dag (d)$ are the creation (annihilation) operators of leads and QD, respectively. $\epsilon_n$ is the energy level for the QD and $\epsilon_{k\alpha}$ is the energy levels of the lead $\alpha (\alpha = L, R)$. $t_{k\alpha n}$ is the coupling constant between two leads and the QD.

To investigate the energy current through the left lead where the measurement is made, we focus on the energy operator of the left lead
\begin{equation}\label{hamL}
H_L = \sum_{k} \epsilon_{kL} c^\dag_{kL} c_{kL}.
\end{equation}
Since we study the behaviour in the transient regime we assume that the bias is applied to the leads at $t=-\infty$ while the leads and QD are disconnected. All the couplings are switched on at $t=0$ giving rise to a transient energy current. The switching of coupling between QD and leads can be done by a quantum point contact that is controlled by a gate voltage. Since the system is disconnect before $t=0$, the initial density matrix of the whole system at time $0$ is the direct product of the subsystems expressed by $\rho(0)=\rho_L\otimes\rho_D\otimes\rho_R$. Similar to the cases of phonon and electron charge transport\cite{W2,T2}, the generating function of transferred energy can be expressed as,
\begin{equation}\label{gf1}
Z(\lambda,t) = \mathrm{Tr}\left[ \rho(0) e^{i\lambda H_{L}(0)} e^{-i\lambda H^h_{L}(t)} \right].
\end{equation}
Here, $H^h_{L}(t)$ denotes the energy operator of the lead $L$ in the Heisenberg picture, which is related to the energy operator in the Schr\"{o}dinger picture $H_{L}(0)$ (Eq.~(\ref{hamL})) by
\begin{equation}
H^h_{L}(t)=U^\dag(t,0)H_{L}(0) U(t,0),
\end{equation}
where $U(t,0)$ is the evolution operator.
%\begin{equation}
%U(t,0) = \mathcal{T} \exp\left[ -\frac{i}{\hbar}\int_{0}^{t} H(t') dt'\right],
%\end{equation}
%with $\mathcal{T}$ the time-ordering operator.

In terms of the modified Hamiltonian $H_\gamma$ given in Eq.~(\ref{mH}), the generating function can be rewritten as,
\begin{equation}\label{gf2}
Z(\lambda,t) =\mathrm{Tr}\left\{ \rho(0) U^\dag_{\lambda/2} (t,0) U_{-\lambda/2} (t,0) \right\},
\end{equation}
where the modified evolution operator is,
\begin{equation}\label{U}
  U_\gamma(t,0) = \mathcal{T} \exp\left[ -\frac{i}{\hbar}\int_{0}^{t} H_\gamma(t') dt'\right],
\end{equation}
with
\begin{eqnarray}\label{mH}
%  H_\gamma &=& e^{i\gamma H_L(0)} H e^{-i\gamma H_L(0)} \nonumber\\
   H_\gamma
  &=& \sum_{k} \Big[ \epsilon_{kL} c^\dag_{kL}(t_\gamma) c_{kL}(t_\gamma) + \epsilon_{kR} c^\dag_{kR} c_{kR} \Big]  \nonumber\\
  && +\sum_n \epsilon_n d^\dag_n d_n  + \sum_{kn}\Big[ \Big( t_{kL n} c^\dag_{kL}(t_\gamma) d_n \nonumber\\
  && + t_{kR n} c^\dag_{kR} d_n \Big)  + \mathrm{h.c.} \Big]. \label{H_m}
\end{eqnarray}
with $t_\gamma=\hbar\gamma$ and $\gamma=\lambda/2$.
In deriving Eq.~(\ref{H_m}), we have used the following relation,
\begin{equation}\label{ck}
  e^{i\gamma H_L} c_L(0) e^{-i\gamma H_L} =\sum_n \frac{\hbar^n \gamma^n}{n!} [\partial^n_t c_{kL}(t)]_{t=0}= c_L(t_\gamma).
\end{equation}

%Since $U^\dag_{\lambda/2}(t,0)$ evolves from $t$ to 0 and $U_{-\lambda/2}(t,0)$ goes from $0$ back to $t$, we can introduce the Keldysh contour to combine  $U^\dag_{\lambda/2} (t,0) U_{-\lambda/2} (t,0)$ in Eq.~(\ref{gf2}), in which the upper and lower branch is represented by $\gamma_{\pm}(t) = \mp (\lambda/2) \theta(t)$, respectively. The generating function can be expressed on the Keldysh contour as \cite{T2},
%\begin{equation}\label{gfK}
%Z(\lambda,t) = \mathrm{Tr}\left\{ \rho(0) \mathcal{T}_c \exp\bigg[ -\frac{i}{\hbar}\int_{c} H_\gamma(\tau) d\tau \bigg] \right\}.
%\end{equation}

%For the convenience of deriving the generating function by the Keldysh formalism, it is useful to use the

Using Grassmann algebra the generating function becomes\cite{T2},
\begin{equation}\label{gfgra}
  Z(\lambda,t) = \int D[\bar{\phi}\phi]e^{iS[\bar{\phi}\phi]},
\end{equation}
where $D[\bar{\phi}\phi] = \Pi_{x\sigma} d\bar{\phi}^\sigma_x d\phi^\sigma_x$ with $x \in k\alpha, n$ and the action $S[\bar{\phi}\phi]$ is given by,
\begin{eqnarray}\label{action}
S[\bar{\phi}\phi] &=& \int_{0}^{t} d\tau \sum_{k\sigma} \sigma \Big[ \bar{\phi}^\sigma_{kL}(\hbar\gamma_\sigma)(i\partial_\tau - \epsilon_{kL})\phi^\sigma_{kL}(\hbar\gamma_\sigma) \nonumber\\
&& + \bar{\phi}^\sigma_{kR}(i\partial_\tau - \epsilon_{kR})\phi^\sigma_{kR} \Big] + \sum_{n \sigma} \sigma \bar{\phi}^\sigma_n (i\partial_\tau - \epsilon_{n})\phi^\sigma_{n} \nonumber\\
&& - \sum_{kn\sigma} \sigma \Big[ t_{kL n} \bar{\phi}^\sigma_{kL}(\hbar\gamma_\sigma) \psi^\sigma_n + t_{kR n} \bar{\phi}^\sigma_{kR} \psi^\sigma_n  + \mathrm{c.c.} \Big], \nonumber\\
\end{eqnarray}
where $\phi$ and $\bar{\phi}$ are the Grassmann variables which are two independent complex numbers \cite{gras} and $\sigma = +,-$ denoting the upper and lower branches of the Keldysh contour, respectively.

After the Keldysh rotation\cite{Kamenev} we rewrite the action in Eq.~(\ref{action}) in a matrix form,
\begin{equation}\label{ac-ma}
  S[\bar{\Psi}\Psi] = \int_{0}^{t} d\tau \int_{0}^{t} d\tau' \bar{\Psi}^T(\tau) M(\tau,\tau') \Psi(\tau'),
\end{equation}
with $\bar{\Psi}^T(\tau) = [\bar{\psi}_{kL}^T(\tau_\gamma),\bar{\psi}_n^T(\tau),\bar{\psi}_{kR}^T(\tau)]$ and
\begin{equation}\label{Mmatrix}
  M = \left(
        \begin{array}{ccc}
          g^{-1}_{kk'L}(\tau_\gamma,\tau'_\gamma) & -t_{kL n'}\delta & 0 \\
          -t^*_{k'L n}\delta & g^{-1}_{nn'}(\tau,\tau') & -t^*_{k'R n}\delta \\
          0 & -t_{kR n'}\delta & g^{-1}_{kk'R}(\tau,\tau') \\
        \end{array}
      \right),
\end{equation}
where $\delta$ is a unit matrix in Keldysh time space.

Using the functional integration of the Gaussian integral for the Grassmann fields, the generating function can be expressed by the Keldysh non-equilibrium Green's function as \cite{T2},
\begin{equation}\label{gf}
  Z(\lambda, t) = \mathrm{det} (G\widetilde{G}^{-1}),
\end{equation}
with
\begin{eqnarray}
G^{-1} &=& g^{-1} - \Sigma_L - \Sigma_R, \label{G1} \\
\widetilde{G}^{-1} &=& g^{-1} - \widetilde{\Sigma}_L - \Sigma_R. \label{G2}
\end{eqnarray}

Here, $g = g_{nn'}(\tau,\tau')$ is the Green's function of the isolated QD in Keldysh space. $\Sigma_R = \sum_{kk'}t^*_{k'R n} g_{kk'R} (\tau,\tau') t_{kR n'}$ is the self-energy of the right lead in the Keldysh space in the time domain. $\widetilde{\Sigma}_L = \sum_{kk'}t^*_{k'L n} g_{kk'L} (\tau_\gamma,\tau'_\gamma) t_{kL n'}$ is the self-energy with the counting field, meaning that the two-time measurement is done in the left lead.  It is easy to find the expression of $\widetilde\Sigma_L$ given by,
\begin{eqnarray}\label{gk}
  \widetilde\Sigma_L(t,t') &=& \left(
                          \begin{array}{cc}
                            \widetilde\Sigma_L^r(t,t') & \widetilde\Sigma_L^k(t,t') \\
                            \widetilde\Sigma_L^{\bar{k}}(t,t') & \widetilde\Sigma_L^a(t,t') \\
                          \end{array}
                        \right),
\end{eqnarray}
where
\begin{eqnarray}\label{tildesig2}
   \widetilde\Sigma^r_L&=&\frac{1}{2} \left( \Sigma^r_L + \Sigma^a_L - \widetilde\Sigma^{<}_L + \widetilde\Sigma^{>}_L\right), \nonumber\\
   \widetilde\Sigma^k_L&=&\frac{1}{2} \left( \Sigma^k_L + \widetilde\Sigma^{<}_L + \widetilde\Sigma^{>}_L\right), \nonumber\\
   \widetilde\Sigma^{\bar{k}}_L&=&\frac{1}{2} \left( \Sigma^k_L - \widetilde\Sigma^{<}_L - \widetilde\Sigma^{>}_L\right),
\end{eqnarray}
with $\Sigma_L^k = 2\Sigma_L^< + \Sigma_L^r - \Sigma_L^a$.

Note that the expression of Eq.~(\ref{gf}) is the same as that of FCS of charge transport in Refs.~\onlinecite{T2} and \onlinecite{ep1}. The difference lies in the expression of $\widetilde\Sigma_L$. For the charge transport the counting field appears as an extra phase while for the energy transport the presence of counting field is to shift the time from $t$ to $t_\gamma$. In Eq.~(\ref{gf}) the generating function is expressed in terms of a determinant in both time and space domains. As a result, the calculation of generating function is very time consuming for realistic systems. To calculate $Z(\lambda, t)$ for fixed $\lambda$ and $t$ for a system with $N$ degrees of freedom, we need to discretize the time $t$ into $N_t$ uniform mesh and calculate the Green's function as a function of both position and time. The computational complexity to evaluate the determinant is of order $N^3 N_t^3$ for each $t$ \cite{foot3}. For this reason, the transient calculation of FCS is limited to a single QD or double QD with $N=1$ or $2$ where the exact solution of time-dependent non-equilibrium Green's function is available. If one wishes to study a realistic system with $N=100$ (for instance), the amount of calculation increases by six orders of magnitude. However, if we are interested in the first a few cumulants, we can first find their expressions using Eqs.(\ref{jth}) and (\ref{gf}) and then calculate them numerically using these expressions rather than calculating cumulant generating function numerically.

Now we examine the limiting cases of generating function defined in Eq.~(\ref{gf}). First of all, we look at the energy current in the transient regime. From Eq.~(\ref{gf}) and using the relation $\ln \det A = \mathrm{Tr} \ln A$, the cumulant generating function is given by
\begin{equation}\label{cgf}
  \ln Z(\lambda, t) = \mathrm{Tr} \ln [I - G( \widetilde{\Sigma}_L - \Sigma_L)],
\end{equation}
According to Eq.~(\ref{jth}), the transient energy current can be expressed as (see derivation in appendix~\ref{a1} and we have set $\hbar = e = 1$),
\begin{equation}\label{1st}
I^{E}_L(t) = 2\mathrm{Re} \int dt' \mathrm{Tr} \big[ G^r(t,t') \breve{\Sigma}^<_L(t',t) + G^<(t,t') \breve{\Sigma}_L^a(t',t) \big],
\end{equation}
where
\begin{equation}\label{sigma}
  \breve{\Sigma}^\chi_L(t',t) = \sum_{k} \epsilon_{kL} \Sigma^\chi_{kL}(t'-t),
\end{equation}
with $\chi = <,a$ and $\Sigma^\chi_{kL}(t'-t)$ being the self-energy of the left lead in the absence of the counting field. This expression of transient energy current agrees with that obtained directly by the Green's function method \cite{Yu}.

We now consider the short time behavior of the generating function. Using the fact that $G( \widetilde{\Sigma}_L - \Sigma_L)$ is of order $t^2$, we find for small t
\begin{equation}
  \ln Z(\lambda, t) = -\mathrm{Tr} [G( \widetilde{\Sigma}_L - \Sigma_L)],
\end{equation}
where the quantities in trace are in Keldysh space. It is straightforward to show
\begin{equation}\label{cgf1}
  \ln Z(\lambda, t) = \mathrm{Tr} [G^<( \widetilde{\Sigma}^>_L - \Sigma^>_L)+G^>( \widetilde{\Sigma}^<_L - \Sigma^<_L)].
\end{equation}
In the limit of weak coupling case for a single quantum dot system with a single level, we can use Green's function of isolated quantum dot to replace $G^<$ and $G^>$ and we find
\begin{eqnarray}\label{short}
\ln Z(\lambda, t)=(n_d-1) M_{L1} + n_d M_{L2},
\end{eqnarray}
where $n_d$ is the initial occupation number of the isolated QD and
\begin{eqnarray}
M_{L1}&=&\int dE A_0(E) (e^{i \alpha(E) \lambda}-1) f_L(E), \\
M_{L2}&=&\int dE  A_0(E) (e^{-i \alpha(E) \lambda}-1) (f_L(E)-1),
\end{eqnarray}
with
\begin{eqnarray}\label{a0}
A_0(E)=\frac{4\Gamma_L(E)}{\pi} \frac{\sin^2[(E-\epsilon_0)t/2]}{(E-\epsilon_0)^2}.
\end{eqnarray}
where $\Gamma_L(E)$ is the linewidth function of the left lead. Here, $\epsilon_0$ is the energy level of the QD and $\alpha(E) = 1$ or $E$ for charge transport and energy transport, respectively. In the wideband limit (WBL) and $\alpha=1$, Eq.~(\ref{short}) recovers the result of Ref.~\onlinecite{ep1} from which a universal behavior of $n$th cumulant of charge transport $C_n$ has been derived that was first demonstrated experimentally in Ref.~\onlinecite{Flindt2}. Note that an important relation holds for charge transport when $n_d=0$, i.e.,
\begin{equation}\label{rel}
(-i)^n \partial^n Z(\lambda,t)/\partial \lambda^n |_{\lambda=0}= x(t),
 \end{equation}
which is independent of $n$ with $n>0$. This allows one to obtain an analytic expression for $C_n$ for charge transport in the short time limit and very weak coupling regime leading to this universal behavior.

For the energy transport with $\alpha=E$, Eq.~(\ref{rel}) does not hold anymore. Although no analytic expression is available for $n$th cumulant of energy transport $C_n$, some asymptotic behavior can be derived. Assuming $n_d=0$, i.e., initially there is no electron in the QD. From Eq.~(\ref{short}), it is straightforward to find the expression of $n$th order cumulant of energy transport ($\alpha=E$) at zero temperature
\begin{eqnarray}\label{cumu0}
C_n=\frac{\partial^n \ln Z}{\partial (i\lambda)^n} =  \int_{-\infty}^{\Delta_L} dE  A_0(E) E^n,
\end{eqnarray}
where $\Delta_L$ is the bias voltage of the left lead. Now we use WBL such that $\Gamma(E)$ is a nonzero constant only for $|E|<W$\cite{Urban}. In the regime $W=\Delta_L \gg |\epsilon_0|>0$, we have
\begin{eqnarray}\label{cumu1}
C_{n}=  \int_{-W}^{W} du  A_0(u) (u+\epsilon_0)^{n},
\end{eqnarray}
where $u=E-\epsilon_0$. Since $A_0(u)$ is an even function, this integral can be evaluated in the large W limit. For even $n=2k$, the major contribution comes from $\int du A_0(u) u^{2k}$ which gives
\begin{eqnarray}\label{cumu3}
C_{2k} \sim  a_1 W^{2k-1},
\end{eqnarray}
with $a_1=\frac{4\Gamma_L}{(2k-1)\pi}$, the next order $W^{2k-2}$ term depends on $t$.
For $C_{2k+1}$, it is dominated by $\int du A_0(u) (2k+1) \epsilon_0 u^{2k}$ from which we find
\begin{eqnarray}\label{cumu4}
C_{2k+1} \sim  a_2 W^{2k-1},
\end{eqnarray}
with $a_2=\frac{4\Gamma_L(2k+1)}{(2k-1)\pi} \epsilon_0$. Denoting $F_{n} = \ln(C_{n}/C_1)$, we have $F_{2k} = \ln(a_1/C_1) +(2k-1)\ln W $ and $F_{2k+1} = \ln(a_2/C_1) +(2k-1)\ln W $. This suggests that both $F_{2k}$ and $F_{2k+1}$ depend linearly on $k$ with the same slope but different intercepts. Since the slope depends only on the bandwidth $W$, it is universal. Obviously, the above discussion is qualitative under certain limits, the detailed numerical study on the universal behavior at short time will be presented in the next section.

Now we investigate the long-time behavior of the generating function in the transient regime. When $t$ goes to infinity, the Green's function and self-energy in the time domain become invariants under the time translation \cite{T2}. Therefore, the cumulant generating function in Eq.~(\ref{cgf}) in the long-time limit in the energy space becomes,
\begin{equation}\label{cgflt}
  \ln Z_s (\lambda,t) = t \int \frac{d\omega}{2\pi} \ln \det \big\{ I - G(\omega) [\widetilde{\Sigma}_L(\omega) - \Sigma_L(\omega)]\big\}.
\end{equation}

In the next section, we will give numerical result of FCS of transferred energy in the transient regime. We will study FCS for two systems, single QD and double QD systems. In calculating the generating function numerically from Eq.~(\ref{gf}), the Green's functions defined in Eq.~(\ref{G1}) for an occupied single QD can be expressed as,
\begin{eqnarray}
  g^r(\tau_1, \tau_2) &=& -i\theta(\tau_1-\tau_2) \exp[-i\epsilon(\tau_1-\tau_2)], \label{gr}\\
  g^<(\tau_1, \tau_2) &=& i \exp[-i\epsilon(\tau_1-\tau_2)], \label{gl}
\end{eqnarray}
with the Heaviside step function $\theta(\tau_1-\tau_2)$. For a double QD system, the Green's function with the counting field in Eq.~(\ref{G2}) should be written as,
\begin{equation}\label{Gd}
  \widetilde{G}^{-1} = \left(
                    \begin{array}{cc}
                      g^{-1}_1 - \widetilde{\Sigma}_{L} & -v \\
                      -v^* & g^{-1}_2 - \Sigma_{R} \\
                    \end{array}
                  \right).
\end{equation}
Here, $g^{-1}_1$ and $g^{-1}_2$ are the Green's function for the first and second isolated QD, respectively. $v$ is the coupling constant between two QDs.

In order to calculate the self-energy $\Sigma_{L(R)}$ and $\widetilde{\Sigma}_L$ in Eqs.~(\ref{G1}) and (\ref{G2}), we use the Lorentzian linewidth function to describe the self-energy so that the equilibrium energy dependent self-energy can be written using a finite band width $W$,
\begin{equation}\label{selfenergy}
  {\Sigma}^r_\alpha(\omega) = \frac{\Gamma_\alpha W}{2(\omega + iW)},
\end{equation}
with the linewidth amplitude $\Gamma_\alpha$. This is a special model that allows us to find the Green's function exactly while still going beyond the WBL. Note that one can not tune the bandwidth experimentally. Then the retarded self-energy can be given by,\cite{T2}
\begin{eqnarray}\label{sigmar}
\Sigma^r_\alpha (\tau_1, \tau_2) &=& -\frac{i}{4} \theta(\tau_1-\tau_2) \Gamma_\alpha W e^{-(i\Delta_\alpha + W)(\tau_1 - \tau_2)},
\end{eqnarray}
where $\Delta_\alpha$ is the external bias applied on the lead $\alpha$. In order to calculate the lesser Green's function analytically, we focus on zero temperature. We find $\Sigma^<_\alpha (\tau_1 - \tau_2) = \frac{i}{8} \Gamma W$ for $\tau_1 = \tau_2$, and otherwise \cite{T2}
\begin{eqnarray}\label{sigmal}
  \Sigma^<_\alpha(\tau_1, \tau_2) &=& \frac{i}{8} \Gamma W \bigg\{ - \frac{i}{\pi} e^{-(i\Delta_\alpha - W)\tau} \mathrm{Ei}(-W\tau) \nonumber\\
   &&+ e^{-(i\Delta_\alpha + W) \tau} \Big[ 1+\frac{i}{\pi} \mathrm{Ei}(W\tau) \Big] \bigg\},
\end{eqnarray}
with $\tau = \tau_1 - \tau_2$ and $\mathrm{Ei}(x) = -\int_{-x}^{\infty} \frac{e^{-t}}{t} dt$. Note that the diagonal element of lesser self-energy diverges for large $W$. It has been confirmed in Ref.\onlinecite{joseph} that the transient charge current at WBL can be obtained as follows: calculating transient current as a function of $W$ and then taking large $W$ limit. We have confirmed that transient energy current at WBL can be obtained similarly.

Before we end this section, we mention that the approach presented in this paper is suitable only for non-interacting problems. Under a special situation where electrons couple with a single phonon mode, this type of approach can be generalized (see Ref.~\onlinecite{Urban}).

\section{Numerical results}\label{sec3}
In this section, we first apply our formalism to a single QD system which is assumed to be half occupied at $t=0$. The dependence of cumulants on the occupation will be examined later. The linewidth amplitude in Eq.~(\ref{selfenergy}) is set to be $\Gamma_L = \Gamma_R = \Gamma/2$ and the bandwidth $W$ is also set to be the same for both leads. The energy level of the QD is assumed to be $5\Gamma$ and a bias with amplitude $\Delta_L = 10\Gamma$ is chosen for this system. In the following numerical calculations, we set $e = \hbar =\Gamma = 1$ for simplicity.

\begin{figure}
  \includegraphics[width=3.25in]{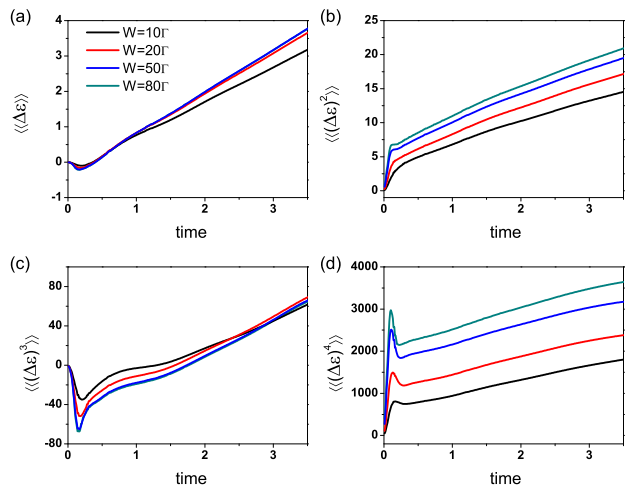}\\
  \caption{(a) 1st, (b) 2rd, (c) 3rd, and (d) 4th cumulants of transferred energy with different bandwidth $W$ in the left lead for a single QD system.}
  \label{fig1}
\end{figure}

Figure~\ref{fig1} shows the 1st to 4th cumulants of transferred energy counted from the initial time $t=0$ in the left lead of the system for different bandwidths $W = 10\Gamma$, $20\Gamma$, $50\Gamma$, and $80\Gamma$. For the 1st and 3rd cumulants, they decrease immediately once the system turns on and increase after reaching a minimum. In this region, the 1st and 3rd cumulants with smaller bandwidths have larger value until the crossover occurs at the time around $0.45$ and $2.5$, as shown in Fig.~\ref{fig1}(a) and \ref{fig1}(c), respectively. After the crossover, situation reverses, i.e., the 1st and 3rd cumulants with the larger bandwidth $W$ becomes smaller than those with smaller $W$ in the self-energy. From Fig.~\ref{fig1}(b) and \ref{fig1}(d) we see that the 2nd and 4th cumulants show a sharp rise first when the system turns on and then increases almost linearly after the transient regime. Roughly speaking the larger the bandwidth $W$, the larger the values of the 2nd and 4th cumulants.

\begin{figure}
  \includegraphics[width=3.25in]{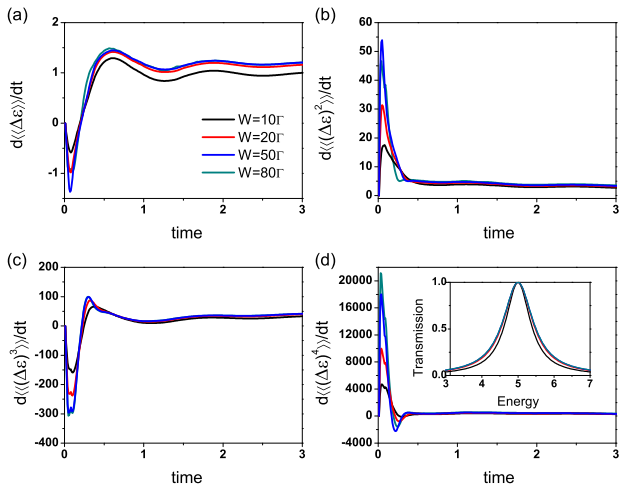}\\
  \caption{Time derivative of (a) 1st, (b) 2rd, (c) 3rd, and (d) 4th cumulants of transferred energy with different bandwidth $W$ in the left lead for a single QD system. Inset: transmission coefficients of the single QD system with different bandwidth.}
  \label{fig2}
\end{figure}

We find that all cumulants of transferred energy increase with small oscillations when the time increases and exhibit linear characteristics at the long time which agrees with the long-time limit of the cumulant generating function in Eq.~(\ref{cgflt}). The small oscillation of cumulants with time can be seen clearly from their derivative with respective of time, as shown in Fig.~\ref{fig2}. Fig.~\ref{fig2}(a) presents the time derivative of the 1st cumulant, namely, the transient energy current of the left lead for different bandwidths $W$. We see that they first drop down exhibiting dips with negative value once the system is connected. After that they increase to maximum values and then decay in an oscillatory fashion. In the long-time limit, they reach the values of dc energy current. We note that the transient energy current behaves similarly to the transient charge current obtained in Ref.~\onlinecite{joseph}. For the time derivative of the 2nd cumulant which is related to the shot noise of the system, it increases immediately exhibiting peaks when the system turns on, and drops down to the long-time limit very quickly with tiny oscillations. The time derivative of the 4th cumulant shows similar behavior to that of the 2nd cummulant besides that it drops to negative values and finally approaches to the positive long-time limit, while that of the 3rd cumulant shows opposite behavior. The short time behavior of $n$th cumulants of energy transport $C_n$ in Fig.~\ref{fig2} can be qualitatively understood from Eq.~(\ref{cumu1}) where $C_n$ depends on $W$ through $\Gamma(E) \sim W^2/(E^2+W^2)$. Hence at short times, a large $W$ gives a large $|C_n|$. The sign of $C_n$ can also be understood from Eq.~(\ref{cumu0}) for $t$ approaching zero. For $n=2k$, we have from Eq.~(\ref{a0}) and (\ref{cumu0})
\begin{eqnarray}
C_{2k}=  \frac{\Gamma_L t^2}{\pi} \int_{-\infty}^{\Delta_L} dE  E^{2k},
\end{eqnarray}
which is positive definite. For $n=2k+1$, since $\int_{-\Delta_L}^{\Delta_L} dE  E^{2k+1}=0$, we have
\begin{eqnarray}
C_{2k+1}=  \frac{\Gamma_L t^2}{\pi} \int_{-\infty}^{-\Delta_L} dE  E^{2k+1},
\end{eqnarray}
which is negative in agreement with the results of Fig.~\ref{fig1} and Fig.~\ref{fig2}.

From the numerical results we find that the frequency of the oscillations is independent of the bandwidth of leads while for the transient energy current, it is found that the time-dependent energy current with larger bandwidth $W$ decays faster than that with smaller $W$. This oscillatory behavior can be understood analytically. For a QD under an upward pulse of bias within the WBL the transient energy current can be expressed in terms of the spectral function $A(\epsilon, t)$ as,
\begin{eqnarray}\label{Iet_wbl}
I^E_L(t) &=& -\Gamma_L \int \frac{d\epsilon}{2\pi} \epsilon  \Big\{  2f_L(\epsilon) \mathrm{Im} [A(\epsilon,t)] \nonumber\\
&&+ \sum_\alpha \Gamma_\alpha f_\alpha(\epsilon) |A(\epsilon,t)|^2 \Big\},
\end{eqnarray}
with
\begin{equation}\label{AA}
  A(\epsilon,t) = \frac{\epsilon-\epsilon_0 + i\Gamma/2 + \Delta_L e^{i(\epsilon-\epsilon_0 + \Delta_L + i\Gamma/2)t}}{(\epsilon-\epsilon_0 + i\Gamma/2)(\epsilon-\epsilon_0 + \Delta_L + i\Gamma/2)}.
\end{equation}
Clearly the oscillatory behavior of transient energy current is due to the oscillatory term $\exp[i(\epsilon-\epsilon_0 + \Delta_L)t - (\Gamma/2)t]$ in the spectral function $A(\epsilon, t)$. We note that the period of oscillation of the transient energy current is only dependent on the energy level $\epsilon_0$ of the QD and the applied bias. In addition, the damping of this oscillation is dominated by life time of the resonant state of the QD which is proportional to $\Gamma$ in the WBL. In our numerical calculation, finite bandwidth $W$ is used which affects the lifetime of the resonant state. To find the influence of $W$ on the lifetime, we investigate transmission coefficient versus energy to look for resonant behavior and find the resonant state (transmission peak) and its lifetime (inverse of peak width). From the inset of Fig.~\ref{fig2}(d), we find that the transmission peak is mediated by the resonant state of the single QD system is broadened by increasing the bandwidth $W$ in self-energy. This shows that the life time of the resonant state is proportional to bandwidth and explains why the transient energy current decays faster for the system with larger bandwidth (see Fig.~\ref{fig2}(a)). From the transmission coefficient shown in the inset of Fig.~\ref{fig2}(d), it is clear that larger $W$ corresponds to a large dc charge or energy current which is consistent with our dc limit of transient energy current.

\begin{figure}
  \includegraphics[width=3.25in]{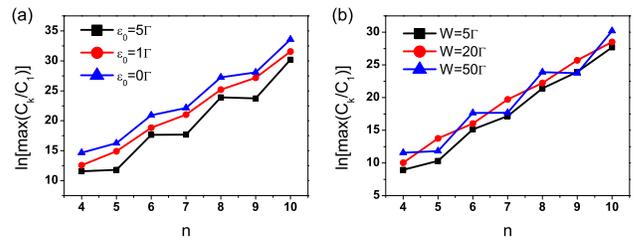}\\
  \caption{The logarithmic plot of the maximum amplitude of the normalized transient energy cumulants $C_n(t)/C_1(t)$ versus $n$ at short times for different system parameters for (a) different $\epsilon_0$ with $W = 50\Gamma$ and (b) different bandwidth $W$ with $\epsilon_0 = 5\Gamma$.}
  \label{figs}
\end{figure}

From the discussion in Section II, we see that there may exist a universal behavior for $C_n$ at short times. In this section, we provide numerical evidence to show that this is indeed the case. Denoting $M_n$ as the maximum amplitude of the normalized transient energy cumulants $C_n(t)/C_1(t)$. In Fig.~\ref{figs}, we show the logarithmic plot of $M_n$ versus $n$ for different system parameters. We see that both $\ln(M_{2k})$ and $\ln(M_{2k+1})$ depend linearly on $k$ with the same slope $\kappa$ but different intercepts. Varying system parameters such as $W$ and $\epsilon_0$ will change the intercepts while the slope remains unchanged. Hence we have $M_{2k} = a_1 e^{\kappa k}$ and $M_{2k+1} = a_2 e^{\kappa k}$ where $a_1$ and $a_2$ are constants. The universal slope $\kappa$ is found to be close to 3.

\begin{figure}
  \includegraphics[width=3.25in]{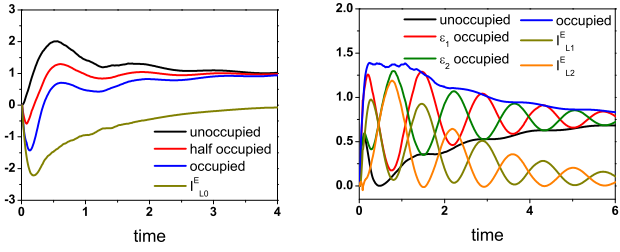}\\
  \caption{(a) Transient energy current of the left lead for a single QD system with $W = 50\Gamma$ which is initially unoccupied (black solid line), half occupied (red solid line), and fully occupied (blue solid line). The dark yellow solid line represents $I^E_{L,in}(t)$ in Eq.~(\ref{oo}) for the fully occupied case. (b) Transient energy current of the left lead for a double QD system with $v = 2\Gamma$ which is initially unoccupied (black solid line), only fully occupied for $\epsilon_1$ (red solid line) or $\epsilon_2$ (green solid line), and fully occupied for both energy levels (blue solid line). The dark yellow and orange solid line represent $I^E_{L,in}(t)$ for the case that only $\epsilon_1$ and $\epsilon_2$ is fully occupied, respectively.}
  \label{figo}
\end{figure}

In order to study the effect of initial occupation number of the single QD on the transient energy current, the time-dependent energy currents calculated by the time-derivative of 1st cumulant with $W = 50\Gamma$ for different initial occupation number are plotted in Fig.~\ref{figo}(a). To understand this behavior,  we note that in the transient regime the lesser Green's function can be expressed by \cite{zwm,leizhang},
\begin{eqnarray}\label{lgt}
G^<(t,t') &=& G^r(t,0)g^<(0,0)G^a(0,t') \nonumber\\
&& + \int_0^t\int_0^t d\tau_1 d\tau_2 G^r(t,\tau_1) \Sigma^<(\tau_1, \tau_2) G^a(\tau_2,t'). \nonumber\\
\end{eqnarray}
Substituting this expression into Eq.~(\ref{1st}) we find the transient energy current consists of two terms $I^E_L(t) = I^E_{L, un}(t) + I^E_{L,in}$ where $I^E_{L, un}(t)$ is the transient energy current for a system which is initially unoccupied while the transient energy current due to the initial occupation is
\begin{equation}\label{oo}
I^E_{L,in}(t) = 2\mathrm{Re} \int dt' \mathrm{Tr} \big[G^r(t,0)g^<(0,0)G^a(0,t') \breve{\Sigma}_L^a(t',t) \big].
\end{equation}
In Fig.~\ref{figo}(a), we plot $I^E_{L,in}(t)$ for the single QD system which is initially full occupied (defined as $I^E_{L0}$), namely, $g^<(0,0) = 1i$. Therefore, the transient energy current for a single dot system with initial occupation of $\alpha$ is $I^E_L(t) = I^E_{L, un}(t) + \alpha I^E_{L0}$. We have checked that three curves (black, red, and blue lines) in Fig.~\ref{figo}(a) indeed satisfy this relation.

As a second example, we consider a double QD system with the Hamiltonian of $\left(
                                                                              \begin{array}{cc}
                                                                                \epsilon_1 & -v \\
                                                                                -v^* & \epsilon_2 \\
                                                                              \end{array}
                                                                            \right)$.
We set the energy levels of the first and second QD to be $\epsilon_1 = 4\Gamma$ and $\epsilon_2 = 6\Gamma$ which connect with the left and right lead, respectively. The bandwidth of the lead and the bias are set to be $W = 10\Gamma$ and $\Delta_L = 10\Gamma$, respectively. The bias is applied to the isolated leads at $t=-\infty$ and the couplings between the leads and QDs are switched on at $t=0$. For the double QD system, we assume that the initial state for $\epsilon_1$ is occupied initially with the lesser Green's function given in Eq.~(\ref{gl}), while $\epsilon_2$ is unoccupied so that its occupation number (proportional to its lesser Green's function) is 0.

\begin{figure}
  \includegraphics[width=3.25in]{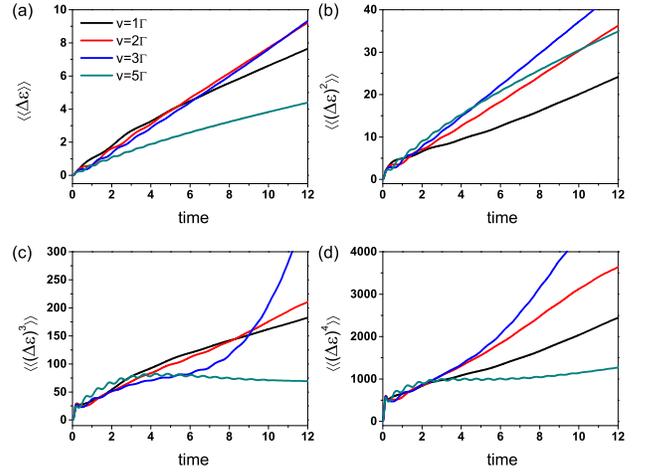}\\
  \caption{(a) 1st, (b) 2rd, (c) 3rd, and (d) 4th cumulants of transferred energy with different coupling constant $v$ in the left lead for a double QD system.}
  \label{fig3}
\end{figure}

Figure~\ref{fig3} presents the 1st to 4th cumulants of transferred energy of electrons counted in the left lead with different coupling constants of $v = 1\Gamma$, $2\Gamma$, $3\Gamma$, and $5\Gamma$ between two QDs. Generally speaking, as the time increases, all calculated cumulants increase with oscillation except for the 3rd cumulant with $v = 5 \Gamma$ which starts to decrease slowly when $t$ exceeds 4.5. The oscillations of the 3rd and 4th cumulants with a large interdot coupling constant $v = 5\Gamma$ decay much more slowly than those of other cases. Similar to the cumulants for the single QD system, the long-time limit of different cumulants for the double QD system is a linear function of time as shown in Eq.~(\ref{cgflt}). It is also found that in the long-time limit, the $n$th cumulant with coupling constant of $v = 3\Gamma$ becomes the largest while $v = \Gamma$ is the smallest compared with those with other coupling constants for all $n$. We also note that the behaivor of cumulants of transferred energy for the double QD system resembles that of cumulants of transferred charge reported in Ref.~\onlinecite{T2}.

\begin{figure}
  \includegraphics[width=3.25in]{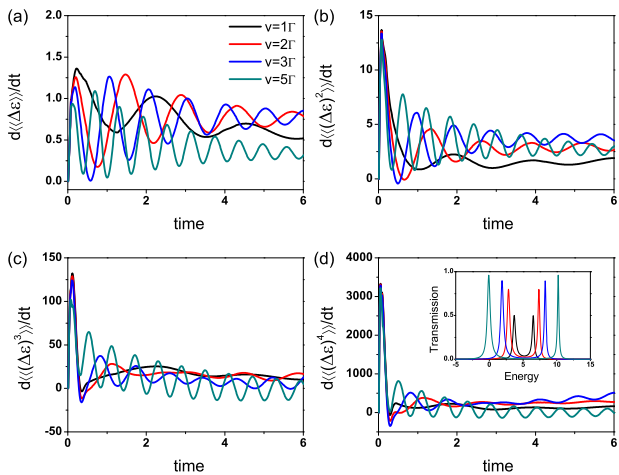}\\
  \caption{Time derivative of (a) 1st, (b) 2rd, (c) 3rd, and (d) 4th cumulants of transferred energy with different coupling constant $v$ in the left lead for a double QD system. Inset: transmission coefficients of the double QD system with different coupling constan.}
  \label{fig4}
\end{figure}

The time derivative of the cumulants for the double QD system are then presented in Fig.~\ref{fig4}. The time derivative of all cumulants show oscillations with their amplitudes decreasing with the increase of time and their frequencies approximately proportional to the coupling constant, especially for large coupling constant. To understand the behavior of oscillation, we note that in the resonance regime the electron oscillates between two QDs with different energy levels and coupling constants, leading to different oscillation frequencies related by the Rabi frequency defined as,
\begin{equation}\label{rabi}
  \omega = \sqrt{(\epsilon_1 - \epsilon_2)^2 + 4|v|^2},
\end{equation}
which is actually the difference between two eigenvalues of the Hamiltonian for the double QD system. In our case, the period of oscillation for the transient energy current is $T = \frac{2\pi}{\omega} = \frac{\pi}{\sqrt{1+ v^2}}$. Therefore, higher frequency is obtained for the time derivative of cumulants plotted in Fig.~\ref{fig4} for system with larger coupling constant. Specifically, for the transient energy current, namely, the time derivative of the 1st cumulant, it is found that for the systems with coupling constants of $v = 1\Gamma$, $2\Gamma$, and $3\Gamma$, larger energy current is obtained for system with larger coupling constant in the long-time limit since the transmission peaks become higher and wider when the coupling constant is increased , as shown in the inset of Fig.~\ref{fig4}(d). However, when the coupling constant is further increased to $5 \Gamma$, the transmission peaks start to shift out of the bias windows [0,10] which dramatically reduces the energy current in the long-time limit. For higher order cumulants, we observe similar behaviors (which can be understood using the augments discussed above): (1). the larger the interdot coupling, the larger the oscillation frequency becomes; (2). in the long-time limit, the value of cumulants increases with $v$ as long as $v<4.5\Gamma$. (3). For $v>4.5\Gamma$, the value of cumulants is the smallest in the long-time limit.

Moreover, the transient energy currents calculated by the time-derivative of 1st cumulant of the double QD with $v = 2\Gamma$ for different initial occupation condition are plotted in Fig.~\ref{figo}(b). Similar to the case of single QD, we also calculate the contribution of transient current from the initial occupation condition by Eq.~(\ref{oo}) for the case that only $\epsilon_1$ ($I^E_{L1}(t)$) and $\epsilon_2$ ($I^E_{L2}(t)$) is fully occupied, respectively, as shown in Fig.~\ref{figo}(b). Therefore, for a double QD system in which the occupation number of $\epsilon_1$ and $\epsilon_2$ are $\alpha$ and $\beta$, respectively, the transient current can be calculated simply by $I^E_L(t) = I^E_{L, un}(t) + \alpha I^E_{L1} + \beta I^E_{L2}$.

\section{Conclusion}\label{sec4}
We have investigated the FCS of transferred energy in the transient regime. Two time measurement scheme was used to derive the generating function of FCS of transferred energy in the transient regime using the Keldysh non-equilibrium Green's function. Our formalism was then applied to both single and double QD systems to study the 1st to 4th order cumulants of transferred energy in the transient regime. Oscillations are observed in the transient energy current for both single and double QD systems. At short times, universal scaling was found for maximum amplitude of normalized cumulant of energy current for the single QD system. For the single QD system, we find that the frequency of oscillation in the transient energy current is independent of the bandwidth of the self-energy while and for the double QD system the frequency is proportional to the coupling constant between two QDs for large coupling constant.

\begin{acknowledgments}
This work was financially supported by the Research Grant Council (Grant No. HKU 705212P), the University Grant Council (Contract No. AoE/P-04/08) of the Government of HKSAR, NSF-China under Grant No. 11374246.
\end{acknowledgments}

\appendix
\section{Transient energy current}\label{a1}
In this appendix, we present the details of obtaining the transient energy current from the cumulant generating function. Accoding Eqs.~(\ref{jth}) and ~(\ref{cgf}), the transient energy current, namely, the 1st cumulant of transfered energy, can be written as ($\hbar = e = 1$),
\begin{eqnarray} \label{a1st}
  I^E_L(t) &=& \frac{\partial \ln Z(\lambda,t)}{i\partial \lambda} \bigg{|} _{\lambda=0} \nonumber\\
  &=& -\mathrm{Tr} \int dt' \bigg[ G(t,t') \frac{\partial \widetilde{\Sigma}_L(t', t) }{i\partial \lambda} \bigg]_{\lambda=0},
\end{eqnarray}
where,
\begin{equation}\label{ag}
G = \left(
      \begin{array}{cc}
        G^r & G^k \\
        0 & G^a \\
      \end{array}
    \right),
\end{equation}
and
\begin{equation}\label{asigma}
  \frac{\partial \widetilde{\Sigma}_L}{\partial \lambda} =  \frac{\partial \left(
                                                                        \begin{array}{cc}
                                                                          -\widetilde{\Sigma}^<_L  + \widetilde{\Sigma}^>_L, & \widetilde{\Sigma}^<_L  + \widetilde{\Sigma}^>_L \\
                                                                          -\widetilde{\Sigma}^<_L  - \widetilde{\Sigma}^>_L, & \widetilde{\Sigma}^<_L  - \widetilde{\Sigma}^>_L \\
                                                                        \end{array}
                                                                      \right) }{2 \partial \lambda},
\end{equation}
from the definition of $\widetilde{\Sigma}_L$ in the Keldysh space in Eq.~(\ref{tildesig2}). For the lesser self-energy $\widetilde{\Sigma}^<_L$ with the counting field, it can be expressed in terms of the lesser green's function of left lead,
\begin{equation}\label{asigma2}
  \widetilde{\Sigma}^<_L(t,t') = \sum_k t^*_{kLn} g^<_{kL} (t-t'-\lambda) t_{kLn},
\end{equation}
where,
\begin{equation}\label{aglead}
  g^<_{kL} (t-t'-\lambda) = i f(\epsilon^0_{kL}) \exp\big[ -i\epsilon_{kL} (t - t' - \lambda)\big],
\end{equation}
with $\epsilon^0_{kL}$ the bare energy levels of the left lead in the absence of external bias.

It is easy to find
\begin{eqnarray}\label{agp}
  \frac{\partial g^<_{kL} (t-t'-\lambda)}{\partial \lambda}\bigg{|} _{\lambda=0}  &=& i\epsilon_{kL} f(\epsilon^0_{kL}) \exp\big[ -i\epsilon_{kL} (t - t')\big] \nonumber\\
  &=& i\epsilon_{kL} g^<_{kL}(t - t'),
\end{eqnarray}
from which we have,
\begin{equation}\label{asp}
   \frac{\partial \widetilde{\Sigma}^<_L(t,t')}{\partial \lambda}\bigg{|} _{\lambda=0} = i\breve{\Sigma}^<_L(t,t'),
\end{equation}
where
\begin{equation}\label{asn}
  \breve{\Sigma}^\chi_L(t,t') = \sum_{k} \epsilon_{kL} \Sigma^\chi_{kL}(t-t'),
\end{equation}
with $\chi = <,a$ and $\Sigma^\chi_{kL}(t'-t)$ being the self-energy of the left lead without the counting field. Similarly, we can obtain,
\begin{equation}\label{asp2}
   \frac{\partial \widetilde{\Sigma}^>_L(t,t')}{\partial \lambda}\bigg{|} _{\lambda=0} = -i\breve{\Sigma}^>_L(t,t'),
\end{equation}
By substituting Eqs.~(\ref{asp}) and (\ref{asp2}) into Eq.~(\ref{asigma}), we have,
\begin{equation}\label{app}
  \frac{\partial \widetilde{\Sigma}_L}{\partial \lambda}\bigg{|} _{\lambda=0} =  \frac{i}{2}\left(
                                                                        \begin{array}{cc}
                                                                          - \breve{\Sigma}^<_L -\breve{\Sigma}^>_L, & \breve{\Sigma}^<_L -\breve{\Sigma}^>_L \\
                                                                          - \breve{\Sigma}^<_L +\breve{\Sigma}^>_L, & \breve{\Sigma}^<_L +\breve{\Sigma}^>_L \\
                                                                        \end{array}
                                                                      \right).
\end{equation}
Therefore, the transient energy current can be obtained from Eq.~(\ref{a1st}),
\begin{eqnarray} \label{ae1}
  I^E_L(t) &=& \frac{1}{2} \int dt' \mathrm{Tr} \Big\{ G^k(t,t')\big[ \breve{\Sigma}^<_L(t',t)  - \breve{\Sigma}^>_L(t',t) \big] \nonumber\\
  && + \big[ G^r(t,t')-G^a(t,t')\big] \big[\breve{\Sigma}^<_L (t',t)+ \breve{\Sigma}^>_L(t',t)\big]  \Big\}. \nonumber\\
\end{eqnarray}
Since $G^k = 2G^< + G^r - G^a$, it can be finally written as,
\begin{eqnarray} \label{ae2}
  I^E_L(t) &=& \int dt' \mathrm{Tr} \Big\{ G^<(t,t')\big[ \breve{\Sigma}^<_L(t',t)  - \breve{\Sigma}^>_L(t',t) \big] \nonumber\\
  && + \big[ G^r(t,t')-G^a(t,t')\big] \breve{\Sigma}^<_L (t',t)  \Big\} \nonumber\\
  &=& 2\mathrm{Re} \int dt' \mathrm{Tr} \big[ G^r(t,t') \breve{\Sigma}^<_L(t',t) + G^<(t,t') \breve{\Sigma}_L^a(t',t) \big]. \nonumber\\
\end{eqnarray}

For the case of heat current, the cumulant generating function for the counting of the heat is found to be,
\begin{equation}\label{cgf_heat}
  \ln Z(\lambda, t) = \textrm{Tr} \ln \big\{ I - G[M(e^{i\sigma_x\mu_L \lambda} - I) + \widetilde{\Sigma}_L - \Sigma_L]\big\},
\end{equation}
with
\begin{equation}\label{M}
  M = \frac{1}{2} \left(
                    \begin{array}{cc}
                      \widetilde{\Sigma}^>_L-\widetilde{\Sigma}^<_L, & \widetilde{\Sigma}^<_L+\widetilde{\Sigma}^>_L \\
                      -\widetilde{\Sigma}^<_L-\widetilde{\Sigma}^>_L, & \widetilde{\Sigma}^<_L-\widetilde{\Sigma}^>_L \\
                    \end{array}
                  \right).
\end{equation}
where ${\tilde \Sigma}_L$ is defined in Eq.~(\ref{tildesig2}).

Therefore, the transient heat current can be expressed as,
\begin{eqnarray} \label{heatI}
  I^h_L(t) &=& \frac{\partial \ln Z(\lambda,t)}{i\partial \lambda} \bigg{|} _{\lambda=0} \nonumber\\
  &=& -\mathrm{Tr} \int dt' \bigg\{ G(t,t') \frac{\partial [M(e^{i\sigma_x\mu_L \lambda} - I) + \widetilde{\Sigma}_L] }{i\partial \lambda} \bigg\}_{\lambda=0} \nonumber\\
  &=& -\frac{\mu_L}{2}[(G^r - G^a)(\Sigma_L^< + \Sigma_L^>) + G^k(\Sigma_L^< - \Sigma_L^>)] \nonumber\\
  && + \frac{1}{2}[(G^r - G^a)(\breve{\Sigma}_L^< + \breve{\Sigma}_L^>) + G^k(\breve{\Sigma}_L^< - \breve{\Sigma}_L^>)] \nonumber\\
  &=& I_L^E(t) - \mu_L I_L(t).
\end{eqnarray}
which is the known result. We also wish to comment that this formalism is valid for heat transport driven by either temperature gradient and bias voltage.


\begin{thebibliography}{40}
\bibitem{Lu} W.~Lu, Z.~Ji, L.~Pfeiffer, K.~W.~West, and A.~J.~Rimberg, Nature \textbf{423}, 422 (2003).
\bibitem{Bylander} J.~Bylander, T.~Duty, and P.~Delsing, Nature \textbf{434}, 361 (2005).
\bibitem{Fuji} T.~Fujisawa, T.~Hayashi, R.~Tomita, Y.~Hirayama, Science \textbf{312}, 1634 (2006).
\bibitem{Belzig} W.~Belzig and Y.~V.~Nazarov, Phys. Rev. Lett. \textbf{87}, 197006 (2001).
\bibitem{Bagrets} D.~A.~Bagrets and Y.~V.~Nazarov, Phys. Rev. B \textbf{67}, 085316 (2003).
\bibitem{Pilgram} S.~Pilgram, A.~N.~Jordan, E.~V.~Sukhorukov, and M.~B\"{u}ttiker, Phys. Rev. Lett. \textbf{90}, 206801 (2003).
\bibitem{Gogolin} A.~O.~Gogolin and A.~Komnik, Phys. Rev. B \textbf{73}, 195301 (2006).
\bibitem{Scho} K.~Sch\"{o}nhammer, Phys. Rev. B \textbf{75}, 205329 (2007).
\bibitem{Saito} K.~Saito and Y.~Utsumi, Phys. Rev. B \textbf{78}, 115429 (2008).
\bibitem{Flindt} C.~Flindt, T.~Novotn\'{y}, A.~Bragggio, M.~Sassetti, and A.-P.~Jauho, Phys. Rev. Lett \textbf{100}, 150601 (2008).
\bibitem{Urban} D.~F.~Urban, R.~Avriller, and A.~Levy Yeyati, Phys. Rev. B \textbf{82}, 121414(R) (2010).
\bibitem{T1} G.-M.~Tang, F.~Xu, and J.~Wang, Phys. Rev. B \textbf{89}, 205310 (2014).
\bibitem{Gust} S.~Gustavsson, R.~Leturcq, B.~Simovi\v{c}, R.~Schleser, T.~Ihn, P.~Studerus, K.~Ensslin, D.~C.~Driscoll, and A.~C.~Gossard, Phys. Rev. Lett. \textbf{96}, 076605 (2006).
\bibitem{Flindt2} C.~Flindt, C.~Fricke, F.~Hohls, T.~Novotn\'{y}, K. Neto\v{c}n\'{y}, T.~Brandes, and R.~J.~Haug, Proc. Natl. Acad. Sci. USA \textbf{106}, 10116 (2009).
\bibitem{Klich} I.~Klich and L.~Levitov, Phys. Rev. Lett. \textbf{102}, 100502 (2009).
\bibitem{Song} H.~Francis Song, C.~Flindt, S.~Rachel, I.~Klich, and K.~Le Hur, Phys. Rev. B \textbf{83}, 161408(R) (2011).
\bibitem{Les} G.~B.~Lesovik, F.~Hassler, and G.~Blatter, Phys. Rev. Lett. \textbf{96}, 106801 (2006).
\bibitem{Ramm} M.~Ramm, T.~Pruttivarasin, and H.~H\"{a}ffner, New J. Phys. \textbf{16}, 063062 (2014).
\bibitem{But} M.~B\"{u}ttiker, Phys. Rev. B \textbf{46}, 12485 (1992).
\bibitem{sivan} U.~Sivan and Y.~Imry, Phys. Rev. B \textbf{33}, 551 (1986).
\bibitem{kearney} M.~J.~Kearney, and P.~N.~Butcher, J. Phys. C. \textbf{21}, L265 (1988).
\bibitem{engine1} B.~Sothmann, R.~S\'{a}nchez, A.~N.~Jordan, and M.~B\"{u}ttiker, Phys. Rev. B \textbf{85}, 205301 (2012).
\bibitem{sanchez1} J.~S.~Lim, R.~L\'{o}pez, and D.~S\'{a}nchez, Phys. Rev. B \textbf{88}, 201304(R) (2013).
\bibitem{sanchez2} M.~F.~Ludovico, J.~S.~Lim, M.~Moskalets, L.~Arrachea, and D.~S\'{a}nchez, Phys. Rev. B \textbf{89}, 161306(R) (2014).
\bibitem{mosk1} F.~Battista, M.~Moskalets, M.~Albert, and P.~Samuelsson, Phys. Rev. Lett. \textbf{110}, 126602 (2013).
\bibitem{mosk2} M.~Moskalets, Phys. Rev. Lett. \textbf{112}, 206801 (2014).
\bibitem{jchen} J.~Chen, M.~ShangGuan, and J.~Wang, New J. Phys. \textbf{17}, 053034 (2015).
\bibitem{Michelini} A.~Cr\'epieux, F.~\v{S}imkovic, B.~Cambon, and F.~Michelini, Phys. Rev. B \textbf{83}, 153417 (2011).
\bibitem{Yu} Z.~Yu, L.~Zhang, Y.~Xing, and J.~Wang, Phys. Rev. B \textbf{90}, 115428 (2014).
\bibitem{SD1} K.~Saito and A.~Dhar, Phys. Rev. Lett. \textbf{99}, 180601 (2007).
\bibitem{SD2} K.~Saito and A.~Dhar, Phys. Rev. E \textbf{83}, 041121 (2011).
\bibitem{C1} A.~A.~Clerk, F.~Marquardt, and J.~G.~E.~Harris, Phys. Rev. Lett. \textbf{104}, 213603 (2010).
\bibitem{C2} A.~A.~Clerk, Phys. Rev. A \textbf{84}, 043824 (2011).
\bibitem{W1} J.-S.~Wang, B.~K.~Agarwalla, and H.~Li, Phys. Rev. B \textbf{84}, 153412 (2011).
\bibitem{W2} B.~K.~Agarwalla, B.~Li, and J.-S.~Wang, Phys. Rev. E \textbf{85}, 051142 (2012).
\bibitem{W3} H.~Li, B.~K.~Agarwalla, and J.-S.~Wang, Phys. Rev. B \textbf{86}, 165425 (2012).
\bibitem{W4} B.~K.~Agarwalla, H.~Li, B.~Li, and J.-S.~Wang, Phys. Rev. E \textbf{89}, 052101 (2014).
\bibitem{foot5} The full counting statistics for time-dependent electronic and energy transport has been studied recently in the presence of electron-phonon coupling \cite{ep1,ep2}.
\bibitem{ep1} R.~S.~Souto, R.~Avriller, R.~C.~Monreal, A.~Martín-Rodero, and A.~L.~Yeyati, Phys. Rev. B \textbf{92}, 125435 (2015).
\bibitem{ep2} B.~K.~Agarwalla, J.-H.~Jiang, and D.~ Segal, Phys. Rev. B \textbf{92}, 245418 (2015).
\bibitem{T2} G.-M.~Tang and J.~Wang, Phys. Rev. B \textbf{90}, 195422 (2014).
\bibitem{gras} J.~W.~Negele and H.~Orland, \textit{Quantum Many Particle Physics} (Westview Press, Boulder, CO, 1998).
\bibitem{Kamenev} A.~Kamenev, \textit{Field Theory of Non-Equilibrium Systems} (Cambridge University Press, Cambridge, 2011).
\bibitem{foot3} Taking advantage of Toeplitz property of the determinant, one can reduce the scaling of $N_t^3$.
\bibitem{joseph} J.~Maciejko, J.~Wang, and H.~Guo, Phys. Rev. B \textbf{74}, 085324 (2006).
\bibitem{zwm} J.~Jin, M.~W.-Y.~Tu, W.-M.~Zhang, and Y.~Yan, New J. Phys. \textbf{12}, 083013 (2010).
\bibitem{leizhang} L.~Zhang, Y.~Xing, and J.~Wang, Phys. Rev. B \textbf{86}, 155438 (2012).
\end{thebibliography}
\end{document}